\documentclass[aps,pre,twocolumn,groupedaddress]{revtex4-2}
\usepackage{lipsum}
\usepackage{graphicx}
\usepackage{amsmath,amssymb}
\usepackage{bm}
\usepackage{dcolumn}
\usepackage{theorem}
\usepackage{wrapfig}
\usepackage{mathtools}
\usepackage{xcolor}
\usepackage{txfonts}
\mathtoolsset{centercolon}
\usepackage{amsmath}
\usepackage{graphicx}
\usepackage{bm}
\usepackage{float}
\usepackage{epstopdf}
\begin{document}
	
	\title{Theoretical model  of the Leidenfrost  temperature}

	\author{Sergey Gavrilyuk\,$^{1}$\ and\ Henri Gouin\,$^{1\,*}$}
	
	%\thanks{}
	\affiliation{
		$^1$Aix Marseille University, CNRS,  
		IUSTI,   UMR 7343, Marseille, France}

	%\date{\today}
	\email[]
	{
	  E-mails: \\  sergey.gavrilyuk@univ-amu.fr\\  \, henri.gouin@univ-amu.fr} 

\begin{abstract}
  The \textit{Leidenfrost effect} is a phenomenon in which a liquid, poured onto a  glowing surface  significantly hotter than the liquid's boiling point, produces a layer of vapor that prevents the liquid from rapid evaporation. Rather than making physical contact,  a  drop of water levitates above the surface.
  
  \noindent The temperature above which the phenomenon occurs is called the \textit{Leidenfrost  temperature}.  The reason for the existence of the  Leidenfrost  temperature, which is  much higher than the boiling point of the liquid, is  not  fully understood and predicted. For water we prove  that the Leidenfrost  temperature corresponds to a bifurcation in the solutions of   equations describing   evaporation  of a nonequilibrium liquid--vapor interface. For water, the theoretical values of obtained Leidenfrost  temperature, and that of the liquid--vapor interface which  is smaller than the boiling point of liquid, fit   the experimental  results  found in the literature.


\end{abstract}
  
\maketitle

\section{Introduction}

When water is projected onto a moderately heated metal plate, it spreads out, starts to boil and evaporates very quickly. Things are quite different when the metal is incandescent: the water temperature  remains below the boiling temperature, divides into numerous droplets that roll,  bounce and at the end of their life {they either take--off or explode. These phenomena  are  well described in Refs. \cite{Walker,Celestini,Quere,Quere1,Pomeau1,Biance,Graeber,Lyu,Sobac,Bouillant}}. The observations also show that the droplets perform translational and rotational motions.   These movements lead to geometrically beautiful patterns. Photographic and stroboscopic tools were then used to describe the experiments, but the effect can be seen with the naked eye. Such a phenomenon is qualitatively very well described in Refs. \cite{Holter,Ma}. An analytical model of these figures and movements has been proposed in Ref.   \cite{Casal}. 	\\ This \textit{Leidenfrost effect},  also called the  boiling crisis,   was carefully observed in 1756 by the German physician J.  G.  Leidenfrost.  Leidenfrost had well understood the cause of the  \textit{film boiling} phenomenon: there is no contact between the  glowing solid  and water, the liquid evaporates in the vicinity of the solid and levitates on a cushion of steam \cite{Leidenfrost}. 

In 1844, M. Boutigny  had also experimented on himself some curious facts related to the phenomenon   such as plunging his hand in a bath of molten iron without burning himself \cite{Boutigny}. Fiery coal can reach about  $540$ degrees  Celsius; candidates for walking on hot coals must moisten their feet to benefit from the Leidenfrost effect.   At the end of the 19th century, physicists multiplied astonishing experiments like transforming  water into ice by pouring it into a crucible containing
	sulphurous acid and heated to  red hot \cite {Larousse}.
	
	Today, it is no longer these curiosities that are the subject of in-depth studies,   a lot of new activities are rising about the Leidenfrost  phenomenon. Besides the  industrial applications involving high temperature processes, the Leidenfrost effect offers new opportunities in self-propelling drops, in drag reduction, in the frictionless transport, in chemical reactors without borders, in heat engines, etc...\cite{Bernardin,Sadasivan1,Sadasivan2,Wang}.  
 The boiling crisis is often the first step in an explosive process that is generated by the contact of a hot surface and a liquid. If it is well dominated by metallurgists for the hardening of metals, it is not yet the case in other fields where it is the cause of important accidents.
For example, in the oil industry, at the bottom of the distillation towers is oil at a temperature of about 400 degrees Celsius. In these towers, very dry steam is injected at the same temperature.  When, due to a malfunction in the installation, liquid water is injected, the explosion that occurs is so violent that it destroys most of the distillation plates \cite{Delhaye}.
In  nuclear industry, several accidents  were initiated by the phenomenon. In 1961,  for the American  SL--1 reactor at  Idaho State Laboratory, an unexpected lifting of a control bar caused water to be projected over the core onto the vessel which, despite its weight of 13 tons, sheared the pipes to which it was connected and rose about 3 meters. In 1986, the  boiling crisis  phenomenon occurred in Chernobyl, and in 2011 in Fukushima, creating major nuclear accidents. 
The largest terrestrial explosion ever recorded, that of the Krakatoa volcano (in 1883) corresponding to 200 megatons of TNT, is also due to the contact of lava  at high temperature  with   sea water.

These events have given rise to a large number of studies \cite{,Nikolayev,Truskinovsky,Pomeau,Zhaoa,Rein}. 
 Of particular interest is  the Leidenfrost temperature i.e., the temperature above which the phenomenon occurs. It  depends on physico--chemical and mechanical properties of the heated  surface, the liquid type and the ambient conditions \cite{van Limbeek_2017,van Limbeek_2021}.    However, it cannot be said that a theory for a satisfactory prediction of the Leidenfrost temperature has been given. The Leidenfrost effect still retains an essential mystery about the reason for a temperature above which there is the creation of the vapor film. One may wonder why, under normal atmospheric pressure, the creation of the film does not occur at a temperature close to 100 degrees  Celsius, the boiling temperature of water, which creates a large quantity of vapor.   \\

In order to treat the problem as simply as possible, we  consider a thin layer of liquid water on a flat, infinite and horizontal solid surface $W$   at a uniform temperature $T_w$. The surface is ideal: it has no additional  physico-chemical properties. Since the layer is thin, we can neglect the gravitational forces.  A schematic description of such a {\it thought  experiment} is shown in Fig.  1. The liquid water layer $\rm (a)$ is assumed to be separated from the  water vapor layer by a liquid--vapor interface $\rm (i)$. We assume that the liquid water layer is under atmospheric pressure $p_0$, and that the liquid--vapor interface  is at temperature $T_i$. The vapor layer between the liquid--vapor interface $\rm(i)$ and the heated surface $W$ is decomposed into two parts:  an intermediate part $\rm (b)$ where the temperature varies from $T_i$ to $T_w$, and the  part $\rm (c)$  at temperature $T_w$ where the vapor is evacuated  along the solid surface  $W$.  It  has been observed that the boiling crisis is always accompanied by a specific frequency regime    called in the literature ${\it 1}/f-noise$ (see Ref. \cite{Truskinovsky} and references therein). We assume that the  vapor density oscillations immediately appear near   the interface and disappear at the end of the part $\rm (b)$.
 We only need  to model  the phase transition at the interface $\rm (i)$ and   non-isothermal and non-homogeneous one--dimensional vapor flow in  (b).   For  $\rm(b)$,  we  use a phase-field model  \cite{Cahn-Hilliard,Widom}. It allows us to find the  {\it bifurcation temperature} below which the existence of such a configuration is not possible.     \\  Without claiming that our model will  solve all the problems posed by the  boiling crisis, we believe that it can help to understand  the phenomenon by explaining for water the origin of the Leidenfrost  temperature.
 \\

To simplify the presentation of the article, we have separate the paper into six sections and three appendices.
In section II we present the classical van der Waals equation of state  and its adjustment to our problem.   Section III studies the thermo-mechanical van der Waals--Korteweg model across the liquid-vapor interface and in the vapor part of the flow.  Sections IV and V study the  dimensionless governing equations  of one-dimensional flows. In Section VI the numerical calculations of the governing equations  are performed related to experimental data  to obtain the   Leidenfrost temperature value. A conclusion ends this presentation. Some technical details are shown in Appendices A, B and C.

\begin{figure}
	\begin{center}
		\includegraphics[width=8.6cm]{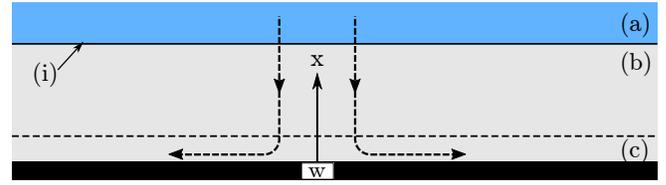}
	\end{center}
	\caption{Sketch of a quasi-one-dimensional transverse fluid flow.  Domain (a) is a thin liquid layer;  (i) is the liquid-vapor discontinuity interface which is a very thin region of few nanometers thickness of vapor having the   temperature $T_i$; domain $(b$) is the non-isothermal part of the vapor flow; the temperature increases  from $T_i$ to $T_w$.  Region  (c) is the  part of the vapor region where the flow is not one--dimensional: the  vapor  escapes  along the solid surface.  The arrows show the flow direction.}
	\label{Fig.1}
\end{figure}
	\section{The van der Waals  equation of state}\label{vdWkey}
We adapt to our problem the simplest model for the description of  equilibrium liquid--vapor interfaces  for water:  the van der Waals equation of state. 
Experimental  values of  physical quantities for water at  the boiling temperature  $T_0=373.15$ Kelvin (corresponding to $ 100$  degrees Celsius) are presented in  International System of Units (SI) (see \cite{HCP}): 
\begin{equation*}
p_0 \approx 101325\ Pa,\,\  v_g \approx 1.673\ m^3/kg, \,\ v_l \approx 0.001043\ m^3/kg,   
\end{equation*}
where $p_0$ is the atmospheric pressure, $v_g$ and $v_l$ are specific volumes of  vapor and   liquid water at phase equilibrium, respectively.  Here and in what follows, we use the SI system. The van der Waals equation of state is 
\begin{equation}
p= \frac{R\,T}{v-b} - \frac{a}{v^2},\label{vdW}
\end{equation}
where $a$, $b$, $R$ are constant,   $v =1/\rho$  is the specific volume, and $\rho$ is the density.  When $v$ is large, the van der Waals equation \eqref{vdW} yields the
	equation of state of perfect gas $p\, v = R\, T$. At a given temperature $T$, one obtains the   chemical potential $\mu$ (defined up to an additive constant),  where $d\mu=v\,dp$: 
\begin{equation}
\mu(v,T)= -{R\,T}\,{\rm{Log}}(v-b) + \frac{R\,T\,b}{v-b}- \frac{2a}{v
}.\label{potchim}
\end{equation}

 The van der Waals equation of state depends on two coefficients $a$ and $b$ and is considered as a good qualitative approximation for the description of equilibrium phase transitions. However, for a given temperature, Maxwell's rule cannot be satisfied because we have to solve three scalar equations  with two unknown scalars $a$ and $b$. Instead of using a more complicated virial form of the equation of state with a large number of temperature dependent coefficients (cf. Refs. \cite{Rocard,Patel}), we adopt a different approach. Rather than the perfect gas constant, we consider a new adaptable parameter.   This simplifies our theoretical approach. To avoid confusion, we write $\mathcal R$ instead of $R$ in Eqs. \eqref{vdW} and \eqref{potchim}. 
 \\
To adapt Eq. \eqref{vdW} to our problem, we calculate the values of $a$, $b$ and $\mathcal R$ to satisfy the mechanical and chemical equilibrium at atmospheric pressure $p_0$.   The  equilibrium Maxwell   conditions of liquid-vapor  interface  are 
\begin{equation}
\left\{
\begin{array}{l}
\displaystyle \quad
\frac{{\mathcal R}\,T_0}{v_g-b} - \frac{a}{v_g^2}=\frac{{\mathcal R}\,T_0}{v_l-b} - \frac{a}{v_l^2}=	p_0, \\ \\
\quad\,\ \displaystyle \mu (v_g,T_0)=\mu (v_l,T_0).  
\end{array}\right.\label{Maxwell rule}
\end{equation}
At  ${T_0=373.15}$\,K\ ($100^\circ$C), we  obtain 
\begin{equation*}
\begin{array}{l}
a\approx 1.52\times 10^3\ m^5\ s^{-2},\,\  b \approx 9.2 \times 10^{-4} \ m^3\ kg^{-1},\,\\ \\  {\mathcal R} \approx 456\ m^{2}\ s^{-2}\ K^{-1} .
\end{array} \label{coefficients vdw_1}
\end{equation*}
 The obtained values of $a$ and $b$ are thus different from those calculated for the thermodynamic critical point \cite{Rocard}. However, the value of ${\mathcal R}$ is close to that of the perfect gas constant which is $R=462  \, m^2s^{-2} K^{-1}$.   
We define a characteristic specific volume $v_0$   of  the vapor phase as:
$$\qquad p_0\,v_0 = {\mathcal R}\, T_0,$$
which gives $$\,\
v_0\approx 1.68 \  m^3\ kg^{-1}.$$ 
Van der Waals' model  is  a qualitatively realistic equilibrium model even far from the boiling point.  Indeed, when we consider   vapor and liquid water  near $160^{\circ}$C,   we obtain from   system \eqref{Maxwell rule} other  values of  coefficients  $a$, $b$ and $\mathcal R$:
\begin{equation*}
\begin{array}{l}
a\approx 1.40\times 10^3\ m^5\ s^{-2},\,\ b \approx 9.3 \times 10^{-4} \ m^3\ kg^{-1},\;\\ \\  {\mathcal R} \approx 447 \ m^{2} \ s^{-2}\ K^{-1} . 
\end{array}\label{coefficients vdw_2}
\end{equation*}
Even the values of $a$, $b$  vary with the temperature, their   effect on the pressure variation  is smaller than $0.5\%$. The variation of   $\mathcal R$ gives an  error in the pressure value  smaller than  $2\%$.  
For the numerical calculations, we use
\begin{equation*}
\begin{array}{l}
a\approx 1.49 \times 10^3\ m^5\ s^{-2},\,\ b \approx 9.2 \times 10^{-4} \ m^3\ kg^{-1},\\ \\ {\mathcal R} \approx 456\ m^{2}\ s^{-2}\ K^{-1}.
\end{array}\label{coefficients vdw_3}
\end{equation*}
 The coefficient $a$ corresponding to the molecular attraction is smaller than the one for classical equilibrium at $100^{\circ}$C.

\section { A continuous theory of capillarity}
   
	We now introduce the second gradient theory of fluids where the internal energy depends on density gradients \cite{Germain,Casal1}. In fact, such a model is a special case of  the Cahn and Hilliard phase field model \cite{Cahn-Hilliard}. It has been developed, in particular, by Rowlinson and Widom \cite{Widom}.

The second gradient theory, conceptually more straightforward than the Laplace
theory, can be used to construct a continuous theory for fluid interfaces. Rowlinson and Widom wrote: \textit{the view that the interfacial  region may be treated as matter in bulk, with a local free-energy density that is that of hypothetically uniform  fluid of composition equal to the local composition, with an additional term arising from the non-uniformity, and that the latter may be approximated by a gradient expansion typically truncated in second order, is then most likely to be successful and perhaps even quantitatively accurate}. The essential difference compared to classical compressible fluids is that  the  specific internal energy depends not only on 
the density $\rho=1/v$,   specific entropy $\eta$, but also  of $\nabla \rho $.
The specific internal   energy $\alpha $ characterizes both the compressibility and
capillarity properties of the fluid. Due to     fluid isotropy, this energy depends only on the norm of density gradient. The simplest expression of the specific  energy   is:
\begin{equation}
\alpha =\varepsilon  (\rho , \eta)+\frac{\lambda}{2\rho }\,\vert\nabla\rho\vert^2, \qquad  \lambda={\rm const}>0.
\label{1}
\end{equation}
Here  $\varepsilon  (\rho , \eta)$ is the classical specific energy and  $\lambda$ is a \textit{capillary coefficient} which is related to the surface tension coefficient:\\
The relation between the surface tension $\gamma$ and coefficient $\lambda$ is given explicitly in  Ref. \cite{Widom} (chapter 3, pages 50--57).   In appropriate CGS units more adapted to capillary phenomena, the value of $\lambda$ is of order $10^{-5}$,  its dimension is $[g]^{-1}\, [ cm]^7\, [s]^{-2}$. It can be calculated  as (formula (3.11) in Ref. \cite{Widom}):
$$\gamma= \int_{-\infty}^{+\infty}\lambda\,\left(\frac{d\rho}{dz}\right)^2\,dz=  \int_{\rho_v}^{\rho_l}\lambda\,\frac{d\rho}{dz}\,d\rho,$$
where  $z$ is the distance in the direction normal to the interface. It can be  approximately written as: 
$$\lambda\approx\frac{\gamma\, h}{({\rho_l}-\rho_v)^2}$$
where $h$ is the thickness of the interface, and $\rho_l$ and $\rho_v$ are the densities of the liquid and vapor, respectively.    \\ Such a gradient density  dependent energy appears      in  the case of large density fluctuations \cite{Zwanzig}.   This is not classical equilibrium thermodynamics of homogeneous states, but thermodynamics of non-homogeneous states.   
 Compared to the classical Laplace theory, the second gradient theory reveals a microstructure of the liquid-vapor interface. Experimental studies of this microstructure have been carried out by the schools of Derjaguin and de Gennes  \cite{Derjaguin,de Gennes}.  
\subsection{Conservative motion}\label{Conservative motion}
For  conservative motion, the van der Waals--Korteweg equations of non-homogeneous capillary fluids  can be derived from the \textit{Hamilton principle of stationary action}    by using the well--known  Lagrangian \cite{Casal0,Gouin1,Gavrilyuk,Gouin2}: 
\begin{equation*}
{\cal L}=\rho\,\left(\frac{ \vert\boldsymbol u\vert^2}{2} - \alpha-  \Omega\right),
\end{equation*}   
where $\boldsymbol u$ is the velocity, $\Omega$ is
the specific potential of external forces, and $\alpha$ is given by Eq. \eqref{1}. 
The usual  constraints are the mass  and entropy conservation laws: 
\begin{equation}
\dfrac{\partial \rho }{\partial t}+\text{div}(\rho\,
\boldsymbol{u})=0,  \label{massconstr}
\end{equation} 
and
\begin{equation}
\dfrac{\partial \rho \eta}{\partial t}+\text{div}(\rho\eta\,
\boldsymbol{u})=0. \label{entropyconstr}
\end{equation} 
We  refer to calculations in Refs. \cite{Germain,Casal1,Gouin} to directly write  the   momentum  equations in the form:
\begin{equation}
\frac{\partial	\rho\, \boldsymbol{u}}{\partial t}+\text{div} \left(\rho\,\boldsymbol{u}\otimes \boldsymbol{u}-\boldsymbol{\sigma}\right) +\rho \, \nabla\,
\Omega = \boldsymbol 0,\label{motion capillary}
\end{equation}
where
\begin{equation*}
\begin{array}{l}\displaystyle
\boldsymbol{\sigma}=-\left(p-\frac{\lambda}{2}\,\vert\nabla\rho\vert^2-\lambda\,
\rho\,\Delta\rho\right)\,\boldsymbol I
-\lambda\,\nabla\,\rho\otimes\nabla\, \rho, \,\\ \\
\displaystyle p=\rho^2\,\frac{\partial \varepsilon(\rho,\eta)}{\partial\rho},
\end{array}
\end{equation*} 
where $I$ is the unit tensor. Due to  a  small thickness of the fluid layer,  gravitational forces  are  neglected.
As a  consequence of Eqs. \eqref{massconstr}, \eqref{entropyconstr} and \eqref{motion capillary}, one obtains the  energy equation :  
\begin{equation}
\frac{\partial e}{\partial t}+ {\rm  div}\left(e\boldsymbol{u}-\boldsymbol{\sigma} \boldsymbol{u}-\lambda\,\frac{d\rho}{dt}\, {\nabla}\,\rho\right) =0, \quad {e=\rho\left(\frac{\vert\mathbf u\vert^2}{2}+\alpha\right)}.
\label{general energy balance}
\end{equation}

In the one-dimensional case, the $x$--axis  is drawn  perpendicular  to the liquid layer and  heated surface   (see Fig. \ref{Fig.1}). The  governing Eq. \eqref{motion capillary} is written as: 
\begin{equation}
\displaystyle
\frac{\partial}{\partial t}(\rho\, u) +\frac{\partial }{\partial x} \left(\rho\, u^2+P\right) =0, \label{motiona}
\end{equation}
where
\begin{equation}
P= p+k,\quad  {\rm with}\quad k=\frac{\lambda}{2}\left(\frac{\partial \rho}{\partial x}\right)^2-\lambda\, \rho\,\frac{\partial^2 \rho}{\partial x^2}.
\label{definition_big_pressure}
\end{equation}
Here $t$ denotes the time, $x$ the space variable perpendicular    to the liquid layer and  glowing surface, $u$  is the corresponding scalar velocity.  Note that $P$ can be considered as a total pressure: it is the sum of the thermodynamic pressure $p$ and \textit{capillary pressure part} $k$. 
	If $k$ is positive (negative), then $P>p$ ($P<p$). 
		In the one-dimensional case, Eq. (7) writes 
		\begin{equation*}
		\frac{ \partial(\rho\, u)} {\partial t}+\frac{ \partial} {\partial {x}}\left(\rho\,u^2+ p(\rho,T)+\frac{\lambda}{2}\left(\frac{\partial \rho}{\partial x}\right)^2-\lambda \,\rho\, \frac{\partial^2 \rho}{\partial x^2} \right)=0
		\end{equation*}
		 At constant temperature $T$, the Gibbs relation becomes   $dp/\rho = d\mu$. Then, using conservation of mass, one obtains 
			\begin{equation*} 
		\displaystyle\rho \left(\frac{\partial u}{\partial t}+u\, \frac{\partial u}{\partial x}\right)+p'(\rho)\frac{\partial \rho}{\partial x} -\lambda \,\rho \, \frac{\partial^3 \rho}{\partial x^3}=0,
		\end{equation*} 
		or  
		\begin{equation*}	\displaystyle\frac{\partial u}{\partial t}+u\, \frac{\partial u}{\partial x}+ \frac{\partial\mu}{\partial x}-\lambda \,  \frac{\partial^3 \rho}{\partial x^3} =0.
		\end{equation*}
		Thus,  
at a given temperature $T$, Eq. \eqref{motiona} admits  the conservation  law:
\begin{equation}
\frac{\partial u}{\partial t}+	\frac{\partial}{\partial x}\left( \frac{u^2}{2}+\mu(\rho, T) -\lambda\, \frac{\partial^2 \rho}{\partial x^2}\right)=0  \label{chemicalpotential}
\end{equation} 
associated with   chemical  potential $\mu$ (a particular case of $\mu$ from Eq. \eqref{potchim} is calculated for the van der Waals equation of state).  

Depending on other additional constraints (isothermal or isobaric processes), we  consider the chemical potential or the specific enthalpy instead of the specific  internal energy  (the details are further explained in Appendices \ref{Isothermal motion} and \ref{Motion at constant pressure}).

\subsection{One-dimensional stationary vapor motion }
 In  the  rest of the paper, one supposes that the consumed liquid is  fed by an external pump  that allows the motion to be steady. We assume one-dimensional flow in the $x$ direction of the  domain (b)   (see Fig. 1).  The viscosity is negligible because the evaporation process is very slow. In the one--dimensional stationary case, Eq. \eqref{massconstr} yields:
\begin{equation}
\rho\, u = q,\qquad q = {\rm const}, \label{mass1}
\end{equation}
where $q$ represents the constant flow rate of the fluid.\\

Equation \eqref{motiona} writes:
\begin{equation}
\displaystyle
\frac{d}{dx} \left(\rho u^2+P\right) =0, \label{motiontrans}
\end{equation}

$\bullet$\quad	Through the liquid--vapor interface,   Eq. \eqref{motiona}  implies the jump condition:
\begin{equation*}
[P+q^2v]=0,
\label{momentum_jump}
\end{equation*}
i.e.
\begin{equation}	P_i-p_0+q^2 (v_{g_i} -v_{l_i})=0,
\label{case b}
\end{equation}
where   the index $i$ refers to the interface:  $v_{l_i}=1/\rho_{l_i}$ ($v_{g_i}=1/\rho_{g_i}$)  the liquid (vapor) specific volume at  interface  (i),  $P_i$ is the total pressure in the vapor phase on the interface, and $p_0$ is the  pressure in the liquid bulk on the interface,  and the square brackets mean  the difference of values across  interface ($i$).  Since the liquid layer is thin, the gravity is not taken into account, thus the  interface liquid thermodynamic   pressure  is  the  atmospheric pressure $p_0$. 	
\\

$\bullet$\quad In  domains (b),  we obtain from Eq. \eqref{motiontrans}:
\begin{equation*}
P_i-p_w + q^2 (v_{g_i}-v_w) = 0,
\end{equation*}
 where the index $w$ corresponds to the heated surface $W$. The vapor   on the boundary between (b)  and  (c) is assumed to be homogeneous of   specific volume $v_w$ and    temperature $T_w$.     Thus, the total pressure is only the thermodynamic pressure part  $p_w$.  The difference with Eq. \eqref{case b} yields:
\begin{equation}
p_w -p_0+q^2( v_w -v_{l_i})= 0. \label{case c}
\end{equation}

$\bullet$\quad  The conservation law \eqref{chemicalpotential} yields the jump relation through the isothermal liquid--vapor interface:  
\begin{equation}
\left[\frac{u^2}{2}+ \mu(\rho,T_i)-\lambda\, \frac{d^2 \rho}{d x^2}  \right] =0,  \label{shock}
\end{equation}
where $\mu(\rho,T_i)$ is defined by Eq. \eqref{potchim}.   
Equation \eqref{shock} can be considered as a dynamical Maxwell rule (see also  Appendix \ref{Isothermal motion}). \\

$\bullet$\quad   The vapor motion in domain (b)   is not isothermal. The viscosity of the vapor phase is negligible, so the equation of motion \eqref{motiontrans} is unchanged. The equation of the energy balance \eqref{general energy balance} in the vapor phase  must take into account   the heat exchange in the vapor region. Such a balance equation is in the  form:     
\begin{equation}
\begin{array}{l}
\displaystyle\left\{(e+P)\,u -\lambda \left(\frac{d\rho}{dx}\right)^2 u\right\}_{\displaystyle\vert \rho_{g_i}, T_i}-  \left\{(e+P)\,u  -\lambda \left(\frac{d\rho}{dx}\right)^2 u\right\}_{\displaystyle\vert \rho_w, T_w} \\ \\ = {\dot Q}_w-{\dot Q}_i. \label{energy cshock} 
\end{array} 
\end{equation}
Compared to Eq. \eqref{general energy balance}, we added in the total energy the balance of  heat fluxes  $\dot Q_w -\dot Q_i$. Also, since the flow volume is fixed, it changes the  expression of $e$ (for proof, see Appendix \ref{Motion at constant pressure}):
\begin{equation*}
e=\rho(u^2/2+{\cal H}) \quad{\rm with} \quad 
 {\cal H}= H+\frac{\lambda}{2\rho}\left(\frac{d\rho}{dx}\right)^2, \quad H=\varepsilon+\,\frac{p_0}{\rho}.
\end{equation*}
Thus,  $\cal{H}$ is the  specific enthalpy of capillary fluid at pressure $p_0$, and $H$ is the enthalpy of a homogeneous fluid at pressure $p_0$. The expression of $P$ is given by Eq. \eqref{definition_big_pressure}. 
In the domain (c) near the surface $W$ the density becomes homogeneous and the  balance law \eqref{energy cshock} becomes 
\begin{equation}
\left\{(e+P)\,u -\lambda \left(\frac{d\rho}{dx}\right)^2 u\right\}_{\displaystyle\vert \rho_{g_i}, T_i}-  \left\{(e+P)\,u \right\}_{\displaystyle\vert \rho_w, T_w}={\dot Q}_w-{\dot Q}_i.
\label{bilan_dissipatif}
\end{equation}
The  vapor density strongly varies near and through the interface. 
\\

At the interface, considered as a discontinuity, 
an extra condition must be added on both sides of the interfacial discontinuity:
\begin{equation}
\frac{d\rho}{d x}  =0\label{additivecond}.
\end{equation}  
 The additional condition (19) called also \textit{Weierstrass-Erdmann condition} is fundamental  in the rest of our paper.  It is    recalled and explained in Appendix \ref{AnnexA}. Also, condition  \eqref{additivecond} is obtained and analyzed in \cite{Gavrilyuk,Gavrilyuk1,Gavrilyuk2}. Physically,  it means  the absence of   microenergy concentration at the  surface of discontinuity. 
Such a condition also appears   when a capillary fluid  is in contact with a surface when the  surface is neither attractive or repulsive \cite{Gouin4,Derjaguin}. \\

The density jump implies $d^2\rho/dx^2<0$, and consequently, due to \eqref{definition_big_pressure}, $k>0$ (this property is  analyzed in   Fig. \ref{8}  upper diagram   of Appendix \ref{oscillations}).

The vapor at temperature $T_w$ is assumed to be homogeneous. Using   relation \eqref{bilan_dissipatif} complemented by Eq. \eqref{additivecond},  we get
\begin{equation*}
\begin{array}{l}
\displaystyle\frac{1}{2}\,\frac{q^3}{\rho_i^2}+p_i\,v_{g_i}\,q-\lambda\,        \,\frac{d^2 \rho_{g_i}}{d x^2}\, q+  H_i\,q+{\dot Q}_i\\ \\ = \displaystyle \frac{1}{2}\,\frac{q^3}{\rho_w^2}+p_w\,v_w\,q +   H_w\, q+{\dot Q}_w.
\end{array}
\end{equation*}
Here indices $``i\, "$ and $``w \,"$ mean   values of variables at the interface and  surface, respectively. In the above relation the second derivative of the  density $\rho_{g_i}$ is -  {\it a  priori} - non-vanishing. \\
Since $k=-\displaystyle\lambda \rho_{g_i}\, \frac{d\rho_{g_i}^2}{dx^2}$\,\   and $P_i = p_i+k$, we get:
\begin{equation}
\frac{1}{2}\, q^2 v_{g_i}^2+ H_i  + P_i v_{g_i} + \frac{\dot Q_i}{q}= \frac{1}{2}\, q^2 v_w^2+ H_w   + p_w v_w + \frac{\dot Q_w}{q}.\label{key}
\end{equation}
Let us  underline that:
\begin{equation*}
p_i=p(v_{g_i},T_i),\quad p_w=p(v_w,T_w).
\end{equation*}
From Eq. \eqref{vdW}, we   have  \cite{Rocard}:
\begin{equation*}
\varepsilon=\int c_v(T)\,dT -\frac{a}{v}, 
\label{energy_vdW}
\end{equation*}
where $c_v(T)$ is the specific heat of water vapor at constant volume.
Equation \eqref{key} implies:
\begin{equation}
\begin{array}{l}
\displaystyle 
{\frac{1}{2}\,q^2\left(v_{g_i}^2-v_{w}^2\right)}+	\int_{T_w}^{T_i}c_v(T) dT  +  2\,k\, v_{g_i}  \\ \\
\displaystyle +\, 2\left(\frac{{\mathcal R}T_iv_{g_i}}{v_{g_i}-b}-\frac{{\mathcal R}T_w v_{w}}{v_{w}-b} \right)-\,\frac{a}{v_{g_i}}\, +\,\frac{a}{v_{w}}\, +\frac{\dot Q_i}{q}-\frac{\dot Q_w}{q}= 0. \label{keyheat}
\end{array}  
\end{equation}
We  approximate  the vapor equation of state  by $p_iv_{g_i}\approx {\mathcal R}T_i$, $p_wv_w\approx {\mathcal R}T_w$, and  introduce  \begin{equation*}
c_p(T)=c_v(T)+{\mathcal R}, \label{CP}
\end{equation*}    
corresponding to the  specific heat at constant pressure  which depends only on  temperature $T$.  We obtain from Eq. \eqref{keyheat}:
\begin{equation}
\begin{array}{l}
{\displaystyle \frac{1}{2}\,q^2\left(v_{g_i}^2-v_{w}^2\right)}+	\displaystyle\int_{T_w}^{T_i}c_p(T)\, dT  +   2\,k\, v_{g_i} \\ \\ + \,{\mathcal R}\left(T_i-T_w\right) \displaystyle +\frac{\dot Q_i}{q}-\frac{\dot Q_w}{q}= 0,   \label{key3}
\end{array} 
\end{equation}
To transform a liquid  into saturated vapor,   we need to supply  latent heat  $L$. At  a given temperature,  and for the van der Waals equation of state, the energy of a saturated vapor is approximately independent  on the pressure. Indeed,  considering the internal energy as a function of $v$ and $T$, one has:  
\begin{equation*} \varepsilon(v_g,T)-\varepsilon(v_{g_s},T)= a\left(\frac{1}{v_{sg}}-\frac{1}{v_g}\right), 
\end{equation*}
where $v_{sg}$ is the specific volume of saturated vapor at pressure $p_{s}$ (index $s$ means {\it saturated}), and $v_{g}$ is the specific volume of vapor at pressure $p_{0}$.  Compared to  the latent heat value, this variation is small even for a large variation of the specific volume of the vapor and we can thus  assume that $\varepsilon(v_g,T)\approx\varepsilon(v_{g_s},T)$. 	
 Let  $L(T)$ be the specific heat of evaporation for  saturated vapor (specific latent heat). One has: 
\begin{equation*}
L(T_i)-L(T_w)=\big(\varepsilon (v_{sg_i},T_i)+p_{si}v_{sg_i}\big) -\big(\varepsilon(v_{sg_w},T_w)+p_{sw}v_{sg_w}\big).
\end{equation*}
The saturated vapor  equation of state being  approximated  as:
\begin{equation*}
p_{si}v_{sg_i}\approx {\mathcal R}T_i,\quad {\rm and } \quad p_{sw}v_{sg_w}\approx {\mathcal R}T_w.
\end{equation*}
Hence, 
\begin{equation*}
L(T_i)-L(T_w)\approx \varepsilon (v_{sg_i},T_i) -\varepsilon(v_{sg_w},T_w)+{\mathcal R}(T_i-T_w).
\end{equation*}
At atmospheric pressure, the water vapor equation of state can  also  be  approximated as: \begin{equation*}
p_0v_{g_i}\approx {\mathcal R}T_i,  
\quad {\rm and } \quad p_0v_{g_w}\approx {\mathcal R}T_w,
\end{equation*}
The specific latent heat  is the amount of heat that must be supplied  to a pure liquid, in our case water, to produce the phase transition. 
	Thus
\begin{equation*}
\frac{\dot Q_i}{q}-\frac{\dot Q_w}{q} = L(T_i)-L(T_w).
\end{equation*} 
This is in agreement with \cite{Gottfried, Biance}. Thus, Eq. \eqref{key3}  yields:
\begin{equation}
\begin{array}{l}\displaystyle
\frac{1}{2}\,q^2\left(v_{g_i}^2-v_{w}^2\right)+\displaystyle\int_{T_w}^{T_i}c_p(T) dT  +   2\,k\, v_{g_i} \\ \\ +\, \displaystyle {\mathcal R}\left(T_i-T_w\right) +L(T_i)-L(T_w)= 0, 
\end{array}  \label{key0}
\end{equation} 

\section{Dimensionless equations of motion}
We now  consider  the  dimensionless form of the governing equations.  The dimensionless variables   are  denoted by the same letters but with an additive \textit{tilde sign}. In particular, the
van der Waals equation of state \eqref{vdW} in  dimensionless form is
\begin{equation}
\tilde p= \frac{\tilde T}{\tilde v-\tilde b} - \frac{\tilde a}{\tilde v^2} \label{vdWad} 
\end{equation}
with	
\begin{equation*}
\tilde a = \frac{a}{p_0\,v_0^2}, \quad  \tilde b = \frac{b}{v_0}, \quad  \tilde T = \frac{T}{T_0}, \quad  \tilde p = \frac{p}{p_0}, \quad \tilde  v = \frac{v}{v_0}, 
\label{coefficients vdw_3}
\end{equation*}
where  
$p_0, T_0$ are defined in Section \ref{vdWkey} and $v_0$ is defined from $p_0\,v_0 = {\mathcal R}\, T_0$.
We also introduce the dimensionless variables associated with   capillary pressure term,  specific volumes and flow rate:
\begin{equation*}
\tilde P_i = \frac{P_i}{p_0},\quad  \tilde k = \frac{k}{p_0},\quad \tilde v_{g_i} = \frac{v_{g_i}}{v_0},\quad \tilde v_{l_i} = \frac{ v_{l_i}}{v_0},\quad \tilde q = \frac{q}{q_0}
\end{equation*}
with $\displaystyle  \,\ q_0= \sqrt{\frac{p_0}{v_0}}.$\\
The equation \eqref{case b} takes the following form:
\begin{equation*}
\tilde P_i -1 + \tilde q^2\, ( \tilde v_{g_i}- \tilde v_{l_i})=0,
\end{equation*} 
The dimensionless flow rate $\tilde q$ is  very small. Indeed, when the solid surface temperature is close to the Leidenfrost temperature, the lifetime of   liquid dramatically increases, typically by a factor of 500 associated with the existence of a vapor layer isolating the liquid bulk. For example, a millimeter liquid layer  on a duralumin surface at $200^\circ$C is observed to float for more than a whole minute  \cite{Biance,Himbert-Biance,Rana}.  So, the  fluid velocity due to the liquid evaporation  is about $1.7\times 10^{-5} \; m \ s^{-1}$, and the flow rate $q\approx 1.7\times10^{-2}\; kg\ m\ s^{-1}$. For $q_0=\displaystyle{\sqrt{\frac{p_0}{v_0}}\approx 245 \;  kg\ m\ s^{-1}}$, one has $\tilde q\approx   7\times 10^{-5}\ll 1$. Consequently,  we  can  neglect $\tilde q^2$ in the dimensionless governing equations.\\

The water vapor at pressure $p_0$ can be considered as a gas and we obtain from Eqs. \eqref{case b} and \eqref{case c}:
\begin{equation*}
p_w\approx P_i \approx p_0,\quad p_0\, v_w \approx {\mathcal R}\, T_w .
\end{equation*}
In dimensionless form we get:
\begin{equation*} 
\tilde v_w\approx \tilde T_w\quad{\rm and}\quad \tilde P_i \approx \tilde p_w \approx 1.  
\end{equation*}
The pressure in  vapor at temperature $T_w$ is also the atmospheric pressure $p_0$. \\

From $p_i\,v_{g_i} = {\mathcal R}\, T_i$, we obtain as a consequence of motion equation  in  domain (i): 
\begin{equation}
\tilde T_i  =\tilde p_i\, \tilde v_{g_i} = (\tilde P_i- \tilde k)\,\tilde v_{g_i} \quad {\rm and} \quad
\tilde P_i\approx 1.
\label{system}
\end{equation}

\noindent{\bf Property}: 
\\
{ \it Since $k >0$ (see Appendix \ref{oscillations}), we must have $\tilde T_i/\tilde v_{g_i} <1$. The limit case corresponds to:} 
\begin{equation}
\tilde T_i = \tilde v_{g_i}.
\label{bifurcation} 
\end{equation}  

\noindent  {We hypothesize that the condition \eqref{bifurcation} defines the value of the  Leidenfrost temperature.  Indeed, as we have already mentioned, the total pressure $P$ is composed of the thermodynamic pressure $p$ and the capillary pressure term $k$. When $k$ is positive, the thermodynamic pressure near the interface will be smaller than the atmospheric pressure in the vapor portion of the fluid. Therefore, the thermodynamic pressure gradient lifts the droplet.  This lifting force can therefore be considered as a kind of Archimedean force (buoyancy force). In the following we will show that  this hypothesis fits with  experimental observations.
	
\section{Dimensionless equations of energy}

\subsection{Liquid--vapor  interface (i)}
The condition \eqref{shock} across the liquid--vapor interface writes:
\begin{equation*}
\frac{1}{2}\, q^2 \, v_{g_i}^2+\mu(v_{g_i},T_i) + k \, v_{g_i} = \frac{1}{2}\, q^2 \, v_{l_i}^2+\mu(v_{l_i},T_i), 
\end{equation*}
and  Eq. \eqref{potchim} yields:
\begin{equation*}
\begin{array}{l}
\displaystyle
\frac{1}{2}\, q^2   v_{g_i}^2+  k   v_{g_i} -{\mathcal R} T_i  \left\{{\rm Log}\left(\frac{v_{g_i}-b}{v_{l_i}-b}\right) -  b \left(\frac{1}{v_{g_i}-b}-\frac{1}{v_{l_i}-b}\right)\right\}\\ \\ \displaystyle +2 a\left(\frac{1}{v_{l_i}}-\frac{1}{v_{g_i}}\right)=0.
\end{array}
\end{equation*}
As proved in Section IV we can neglect  $\tilde q^2$ and  one obtains 
\begin{equation}
\begin{array}{l}
\displaystyle
\tilde k  \, \tilde v_{g_i} -\tilde T_i  \left\{{\rm Log}\left(\frac{\tilde v_{g_i}-\tilde b}{\tilde v_{l_i}-\tilde b}\right) -  \tilde b \left(\frac{1}{\tilde v_{g_i}-\tilde b}-\frac{1}{\tilde v_{l_i}-\tilde b}\right)\right\}\\ \\ \displaystyle +2 \tilde a\left(\frac{1}{\tilde v_{l_i}}-\frac{1}{\tilde v_{g_i}}\right)=0.
\label{Cond interf}
\end{array} 
\end{equation}

\subsection{Non-isothermal vapor-layer  (b)}

For the  specific heat at constant pressure $c_p$, we choose a quadratic model in temperature (see Fig. \ref{Fig.2}):
\begin{equation}
c_p(T) =  K_1  + 2\,K_2\,T + 3\,K_3\,T^2.\label{c_p}
\end{equation}
By integration, we obtain:
\begin{equation*}
\int_{T_w}^{T_i}c_p(T) dT =K_1\left(T_i-T_w\right)+ K_2\left(T_i^2-T_w^2\right)+K_3\left(T_i^3-T_w^3\right).
\end{equation*}
This is the custom to consider locally a linear approximation for $L(T)$ \cite{Lemmon}:  
\begin{equation}
L(T) =   L_0 + L_1 \,T  \qquad {\rm where}\quad L_1< 0. \label{L(T)}
\end{equation}
With  Eqs. \eqref{c_p} and \eqref{L(T)}, Eq. \eqref{key0} becomes:
\begin{equation}
\begin{array}{c}
\frac{1}{2}\,q^2\left(v_{g_i}^2-v_{w}^2\right)+	 2\,k\, v_{g_i}+K_1\left(T_i-T_w\right)+ K_2\left(T_i^2-T_w^2\right) \\ \\
+K_3\left(T_i^3-T_w^3\right) +\,  {\mathcal R}\left(T_i-T_w\right) + L_1\left(T_i-T_w\right)= 0.   \label{cp_integrated}
\end{array}
\end{equation}
Neglecting terms associated with $\tilde q^2$, dimensionless form  of Eq. \eqref{cp_integrated} writes:
\begin{equation}
\begin{array}{c}
2\,\tilde k\, \tilde v_{g_i}+\tilde  K_1\left(\tilde T_i-\tilde T_w\right)+ \tilde  K_2\left(\tilde  T_i^2-\tilde  T_w^2\right)+\tilde K_3\left(\tilde T_i^3-\tilde T_w^3\right) \\ \\
+\;  \left(\tilde T_i-\tilde T_w\right) + \tilde L_1\left(\tilde T_i-\tilde T_w\right)= 0,   \label{key4}
\end{array}
\end{equation}
where
\begin{equation*}
\tilde K_1 = \frac{K_1}{\mathcal R}, \quad \tilde {K_2} = \frac{K_2\,T_0}{\mathcal R}\quad \tilde {K_3} = \frac{K_3\,T_0^2}{\mathcal R},\quad \tilde L_1 = \frac{L_1}{\mathcal R}\,.\label{Tilde}
\end{equation*}
\subsection{Consequences}
In dimensionless form, Eq. \eqref{vdWad}  writes: 
\begin{equation}
\left(\tilde v_{l_i}-\tilde b\right)\,\tilde v_{l_i}^2 -\tilde T_i \, \tilde v_{l_i}^2+   \left(\tilde v_{l_i}-\tilde b\right)\,  \tilde a= 0.\label{key5}
\end{equation}
Using Eqs. \eqref{system}, one  obtains: 
\begin{equation}
\tilde k\; \tilde v_{g_i}=\tilde v_{g_i} -\tilde T_i. \label{key6}
\end{equation}
{Taking into account  Eqs. \eqref{Cond interf}, \eqref{key4} and \eqref{key5}, and by using relation \eqref{key6}, the   system allowing to solve our problem is:
	
	\begin{equation}
	\left\{
	\begin{array}{l}
	\displaystyle 
	\tilde v_{g_i}-\tilde T_i  -\tilde T_i  \left\{{\rm Log}\left(\frac{\tilde v_{g_i}-\tilde b}{\tilde v_{l_i}-\tilde b}\right) -  \tilde b \left(\frac{1}{\tilde v_{g_i}-\tilde b}-\frac{1}{\tilde v_{l_i}-\tilde b}\right)\right\}\\ \\
	\quad\displaystyle +\,2 \,\tilde a\left(\frac{1}{\tilde v_{l_i}}-\frac{1}{\tilde v_{g_i}}\right) =0 ,\\ \\
	\displaystyle 2\,\left(\tilde v_{g_i}-\tilde T_i\right) +\tilde  {K_1}\left(\tilde T_i-\tilde T_w\right)+ \tilde  {K_2}\left(\tilde  T_i^2-\tilde  T_w^2\right) \\ \\
	\quad\displaystyle +\,\tilde {K_3}\left(\tilde T_i^3-\tilde T_w^3\right) + \;\left(\tilde T_i-\tilde T_w\right) + \tilde L_1\left(\tilde T_i-\tilde T_w\right)= 0  , \\ \\
	\displaystyle \tilde T_i- \left(\tilde v_{l_i}-\tilde b\right)\,  \left(\frac{\tilde a}{\tilde v^2_{l_i}}-1\right)=0.\\ 
	\end{array}\right.\label{systeme2}
	\end{equation}}

System \eqref{systeme2} is   a system of three equations relatively to unknowns $\tilde v_{g_i},  \tilde v_{l_i},\tilde T_i$.

\section{Numerical study}

\subsection{Values of specific isobaric capacities of water vapor}

The table, giving the values of specific isobaric capacities for  water vapor, is taken from Refs.  \cite{HCP,Lemmon}.
\begin{widetext}
	
\begin{table}[h]
	{\centering{$%
			\begin{tabular}{|c|c|c|c|c|c|c|c|c|c|c|}
			\hline	
			\hline \multicolumn{1}{|c|}{$ 
				T$  degrees  Celsius} &  90$^{\circ}$C  &  100$^{\circ}$C &  120$^{\circ}$C &   140$^{\circ}$C&
			160$^{\circ}$C&  	180$^{\circ}$C &  200$^{\circ}$C 
			\\ \hline
			\multicolumn{1}{|c|}{ $ 
				T  \ {\rm Kelvin}$} &363 K &373 K&393 K&413 K&433 K&453 K&473 K
			\\ \hline
			\multicolumn{1}{|c|}{ $c_p$} &2042.9& 2080 & 2177 &
			2310.9
			&2488.3 &2712.9&2989.5\\
			\hline \hline
			\end{tabular} 
			$ \vskip 0.1cm }  \caption{Isobaric heat capacity of water vapor is expressed in $J\ kg^{-1}\ K^{-1}$. The different temperature values are given together in degrees   Celsius and Kelvin.} \label{TableKeyX}}  
\end{table}
\begin{table}[h] 
	{\centering{$%
			\begin{tabular}{|c|c|c|c|c|c|c|c|c|c|c|c|c|c|}
			\hline
			\hline \multicolumn{1}{|c|}{ $ 
				T$  degrees  Celsius}  &  90$^{\circ}$C    &  100$^{\circ}$ C &  110$^{\circ}$C  &  120$^{\circ}$C &  130$^{\circ}$C &   140$^{\circ}$C&
			150$^{\circ}$C&  160$^{\circ}$C & 	170$^{\circ}$C  & 	180$^{\circ}$C & 	190$^{\circ}$C& 	200$^{\circ}$C\\
			\hline \multicolumn{1}{|c|}{$ 
				T \ {\rm Kelvin}$}  &  363$^{\circ}$K    &  373$^{\circ}$ K &  383$^{\circ}$K  &  393$^{\circ}$K &  403$^{\circ}$K &   413$^{\circ}$K&
			423$^{\circ}$K&  433$^{\circ}$K & 	443$^{\circ}$K  & 	453$^{\circ}$K & 	463$^{\circ}$K& 	473$^{\circ}$K
			\\  \hline  
			\multicolumn{1}{|c|}{$L $}& 2283.3& 2256.4 & 2229.6& 2202.1 & 2173.7 &
			2144.3
			&2113.7 &2082.0 & 2048.8 &2014.2&1977.9&1939.7 \\
			\hline \hline
			\end{tabular}
			$ \vskip 0.1cm } \caption{ The latent heat of liquid water to be transformed into vapor is expressed in  $kJ\; kg^{-1}$. The different temperature values are given both in degrees Celsius and Kelvin.}} \label{Table}
\end{table}

\begin{table}[h]
	{\centering{$%
			\begin{tabular}{|c|c|c|c|c|c|c|c|c|c|c|c|c|}
			\hline
			\hline 
			\multicolumn{1}{|c|}{  $T_w $ 
				degrees  Celsius} & 93.5$^{\circ}$C & 100$^{\circ}$C &
			119$^{\circ}$C
			&137$^{\circ}$C &156$^{\circ}$C& 175$^{\circ}$C& 193$^{\circ}$C& 212$^{\circ}$C\\
			\hline
			\multicolumn{1}{|c|}{ $ 
				\tilde T_w\,$} &    0.983 & 1  &   1.05 &
			1.10&  	1.15& 1.20 & 1.25 & 1.30
			\\ \hline
			\multicolumn{1}{|c|}{ $\tilde v_{g_i}$} & 0.983 &0.977&
			0.965
			&0.958&0.960&0.971&0.994&1.032
			
			\\
			\hline
			\multicolumn{1}{|c|}{ $\tilde T_i$} & 0.983 & 0.983 &
			0.983
			&0.983
			&0.983&0.983&0.983&0.983  \\
			\hline
			\multicolumn{1}{|c|}{ $\tilde v_{l_i}$} & $0.000621$&  $0.000621$&
			$0.000621$
			& $0.000621$& $0.000621$& $0.000621$ & $0.000621$& $0.000621$ \\
			\hline \hline
			\end{tabular}
			$ \vskip 0.1cm }} \caption{ Calculations  of $\tilde v_{g_i}$, $\tilde T_i$, $\tilde v_{l_i}$ as a function of $T_w$  by using the first linear approximation  \eqref{graph1}.} \label{TableKeyZ1}
\end{table}

\begin{table}[h]
	{\centering{$%
			\begin{tabular}{|c|c|c|c|c|c|c|c|c|c|c|c|c|c|}
			\hline
			\hline 
			\multicolumn{1}{|c|}{$T_w$ degrees Celsius} & 93.5$^{\circ}$C & 100$^{\circ}$C &
			119$^{\circ}$
			&137$^{\circ}$C &156$^{\circ}$C& 175$^{\circ}$C& 193$^{\circ}$C& 212$^{\circ}$C\\
			\hline
			\multicolumn{1}{|c|}{$ 
				\tilde T_w\,$} &    0.983 & 1  &   1.05 &
			1.10&  	1.15& 1.20 & 1.25&1.30
			\\ \hline
			\multicolumn{1}{|c|}{ $\tilde v_{g_i}$} & 0.983 &0.975&
			0.958
			&0.946&0.941&0.947&0.965&0.997 \\
			\hline
			\multicolumn{1}{|c|}{ $\tilde T_i$} & 0.983 & 0.983 &
			0.983
			&0.983
			&0.983&0.983&0.983&0.983 \\
			\hline
			\multicolumn{1}{|c|}{ $\tilde v_{l_i}$} & $0.000621$&  $0.000621$&
			$0.000621$
			& $0.000621$& $0.000621$& $0.000621$ & $0.000621$& $0.000621$ \\
			\hline \hline
			\end{tabular}
			$ \vskip 0.1cm }} \caption{Calculations  of $\tilde v_{g_i}$, $\tilde T_i$, $\tilde v_{l_i}$ as a function of  $T_w$  by using  the second linear approximation \eqref{graph2}  \label{TableKeyZ2}.}
\end{table}
\end{widetext}
 The following quadratic relation  is used linking the  heat capacity at constant pressure  (in  $J\; kg^{-1}\; K^{-1}$) as a function of the  temperature $T$ expressed in   Kelvin:
\begin{equation}
c_p(T)=   
8329+37.13\, T-   0.05460\, T^2\label{Cp_experiments}.
\end{equation} 
Then 
\begin{equation*}
\int_{T_w}^{T_i}c_p(T)\, dT ={K_1}\left(T_i-T_w\right)+ {K_2}\left(T_i^2-T_w^2\right)+{K_3}\left(T_i^3-T_w^3\right),
\end{equation*}
with
\begin{equation*}
{K_1}=8329,\quad {K_2}= 18.56,\quad {K_3}=-0.01820. 
\end{equation*}
Here and in the following, we  do not indicate SI--dimensions of $K_j$ coefficients {$j \in  \{1, 2, 3\}$}.  
The experimental values of $c_p$ are given in Table  \ref{TableKeyX}. The corresponding approximation \eqref{Cp_experiments} is shown in  Fig. \ref{Fig.2}.  We see that relation \eqref{Cp_experiments} fits perfectly with experiment values.
\begin{figure}
	\begin{center}
		\includegraphics[width=\linewidth]{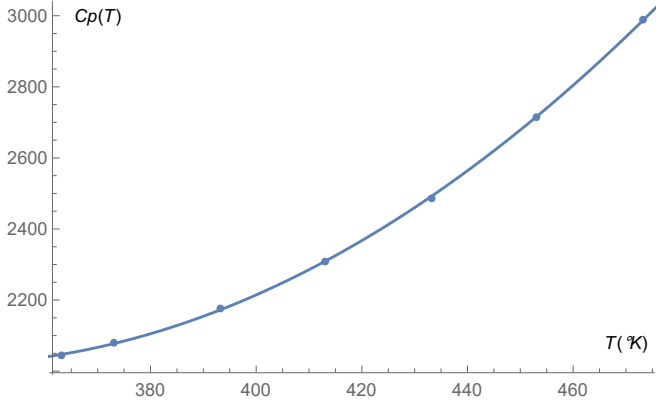}
	\end{center}
	\caption{Graph associated with experimental Table \ref{TableKeyX} and Eq. \eqref{Cp_experiments}. The $x$ axis indicates  the Kelvin temperature, and the $y$ axis indicates for water the corresponding  isobaric heat capacity $c_p$ expressed in $J\; kg^{-1}\; K^{-1}$. The dots represent  $c_p$ values coming from experimental Table \ref{TableKeyX}.} \label{Fig.2}
\end{figure}
\begin{figure}
	\begin{center}
		\includegraphics[width=\linewidth]{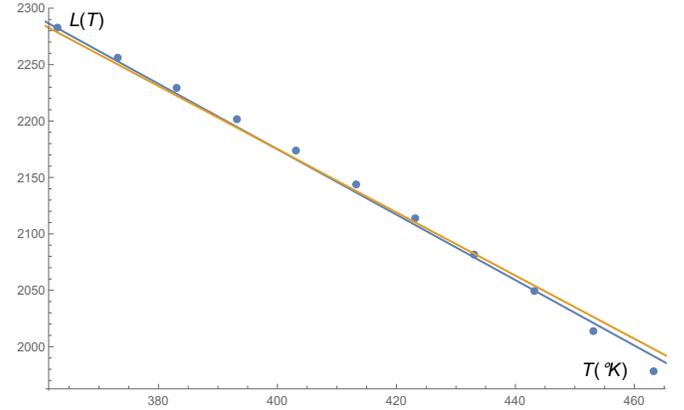}	
		\caption{The linear approximations  of the latent heat of water in  $kJ\; kg^{-1}$   expressed by Eq. \eqref{graph1} (in yellow) and Eq. \eqref{graph2} (in blue) are shown as functions of the Kelvin  temperature.   The dots represent the values of $L(T)$ from  Table II.}  \label{Fig.3}
	\end{center}
\end{figure}

\subsection{Values of latent heat of vaporization for water}

The table  giving the values of latent heat of vaporization for water as a function of temperature is taken from  \cite{HCP,Lemmon}.  
\\
Usually, a local linear approximation of the latent heat $L(T)$  in $kJ \;  kg^{-1}$ (kilojoule per  kilogram) is used  as a function of  temperature $T$ expressed  in  Kelvin.  We consider below two very close approximations of $L(T)$ to understand how the results obtained are sensitive to the values of the latent heat in the numerical calculations.   Indeed, the data shown in Table II correspond to static measurements. In dynamics, the  static latent heat is only a rough approximation: we do not take into account  the heat radiation,  physicochemical  state and geometry of the heating surface, non-equilibrium process of evaporation, etc. 
\begin{itemize}
	\item First linear approximation:
	\begin{equation}
	L(T)= 3295 -  2.800\, T\label{graph1}
	\end{equation}
	\item Second linear  approximation:
	\begin{equation}
	L(T)= 3385 - 2.900 \,T\label{graph2}
	\end{equation}
\end{itemize}

These  two close approximations are shown in  Fig. \ref{Fig.3}.  \\
What matters is the difference of the latent heats $L(T_i)$ and $L(T_w)$. Hence, only the slope in  $T$ is relevant. As we will see, the variation of $3\%$ of  slopes  between Eqs. \eqref{graph1} and \eqref{graph2}, implies a sensible  variation of the Leidenfrost temperature.

 \begin{widetext}
 	
\begin{figure}
	\begin{center}
		\includegraphics[width=11cm]{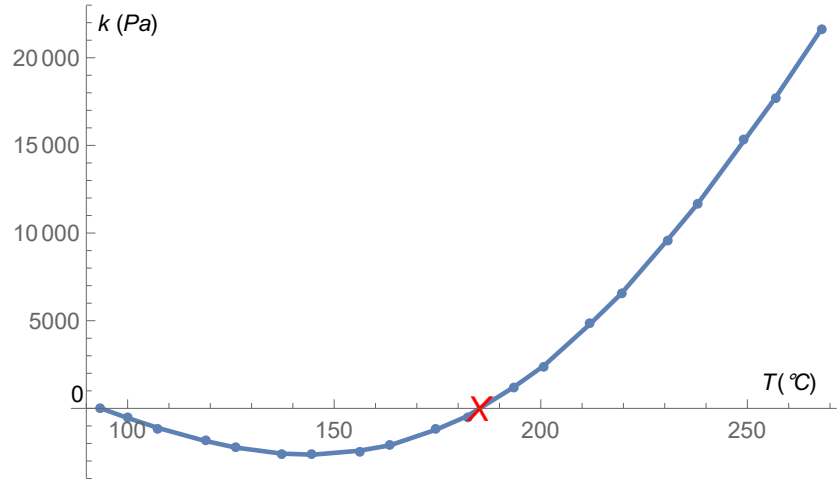}
	\end{center}
	\caption{Graphs associated with the $k$ in  
			$Pa$  as a function of   $T^{\circ}$C   in the \textit{first  linear  approximation} \eqref{graph1}. The $x$ axis is associated with the Celsius temperature and the $y$ axis with the pressure $k$ expressed in Pascal. The dots represented  $k$ values calculated with the software   \textit{Mathematica}$^{TM}$.   In this case, the Leidenfrost temperature  highlighted by a red cross is $T_L\approx 185^ {\circ}$ C.} \label{Fig.4}
\end{figure} 
\begin{figure}
	\begin{center}
		\includegraphics[width=11cm]{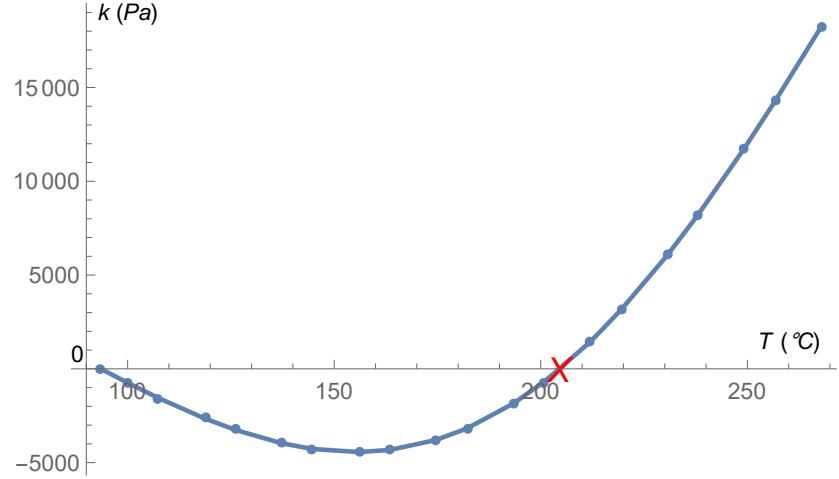}
	\end{center}
	\caption{Graphs associated with the $k$ in  
		$Pa$  as a function of  $T^ {\circ}$C  in the \textit{second linear  approximation} \eqref{graph2}. The $x$ axis is associated with the Celsius temperature and the $y$ axis with the pressure $k$ expressed in Pascal. The dots represented  $k$ values calculated with the software Mathematica$^{TM}$.   In this case, the Leidenfrost temperature        highlighted by a red cross  is $T_L\approx 204^ {\circ}$C.} \label{Fig.5}
\end{figure} 
\end{widetext}
\subsection{Calculations  for water of Leidenfrost and interface temperatures}

 To show the sensitivity of  results to the choice of   model parameters. We have provided two close approximations of the latent heat of evaporation to reveal the sensitivity of the results with respect to these parameters. 
 
For the first linear approximation \eqref{graph1} the corresponding Table \ref{TableKeyZ1} is formed. The condition  $\tilde v_{g_i}=\tilde T_i$ corresponds to the fact that $k$ changes its sign. The value of $T_w$ associated with
bifurcation 	\eqref{bifurcation} is our definition of the Leidenfrost temperature which will be denoted by $T_L$. From   Table \ref{TableKeyZ1} one can see that $\tilde T_i>\tilde v_{g_i}$ at $\tilde T_w=1.20$ but $\tilde T_i<\tilde v_{g_i}$ at $\tilde T_w=1.25$. At $\tilde T_w=\tilde T_L\approx 1.23$ one has $\tilde T_w=\tilde v_{g_i}$.  This critical value is  the Leidenfrost temperature $\tilde T_L$. 	In this case,   $T_L\approx 185^ {\circ}$C.
\\
For  $T_w < T_L$ ($ k<0$)  the liquid  film  sticks to the solid surface by causing the nucleate boiling.
For $T_w > T_L$ ($k>0$) the vapor film exists.  In Fig. \ref{Fig.4}, we represent  the value of $k$ as a function of $T_w$ in degrees Celsius.
\\
For  the second linear approximation \eqref{graph2} the corresponding Table \ref{TableKeyZ2} is formed.  The results are similar but  the associated temperature corresponds to $\tilde T_L\approx 1.28$ i.e. $T_L\approx 204^ {\circ}$C. On Fig. \ref{Fig.5}, we represent  the value of $k$ as a function of $T_w$.  
\\
The  two approximations give  noticeably different temperatures $T_L$  (i.e. the variation of 3\% of the slope of $L(T)$   implies the variation of 10\% on $T_L$).    
\\
Let us note that when $\tilde T_i$ is eliminated from the third equation of Eq. \eqref{systeme2}, only two equations for $v_{l_i}$ and  $v_{g_i}$ have to be solved. We show in Fig. \ref{Fig.6} the intersection of the two corresponding curves for the the first linear approximation \eqref{graph1} and for the value of  $\tilde T_w= 1.23$ corresponding to $T_w= 185^\circ$C. \\

In the literature, a wide range of values of Leidenfrost’s temperature was measured \cite{Bernardin}. The dispersion of values is related to the variation of   experimental conditions (atmospheric conditions, deposition technics, drop size, thermal properties of the substrate, physico-chemical properties of the substrate surfaces (surface energy and roughness),  method to characterize the transition, …).  Depending on the characteristics of the surface, the Leidenfrost temperature can be higher than $T_L \approx 204^\circ$C.
We are not able to account for the various experimental conditions and have considered a flat, highly conductive solid substrate. However, the model provides a correct order of magnitude for the Leidenfrost temperature.
\\
Moreover, it seems that, in the case of water, a minimum Leidenfrost temperature of about $
	150^\circ$C is observed for a liquid film on a flat, highly thermally conductive solid substrate \cite{Liang,Lv}. Nevertheless, it must be pointed out that experiments with ethanol drops on an oil basin can lead to a special Leidenfrost effect for a superheat as low as $T_L-T_0 = 1^\circ$C relative to the boiling temperature. However, this technical feat has never been observed on a solid substrate \cite{Maquet}.
\\

Another important observation results in the  computation of temperature $T_i$. Already Boutigny discovered that this temperature is lower than $100^{\circ}$ C  \cite{Boutigny}.   Experimental data  predict  a  temperature of liquid bulk near the interface between 92$^\circ$ and $97^\circ$ C   \cite {Tokugawa, Bouillant}. The temperature in the liquid bulk far from the interface  depends on the shape of the drop and is linked to heat exchanges with the external environment.    
In our model, the  obtained temperature of interface is $T_i \approx 93.5^{\circ}$C   corresponding  to $\tilde T_i \approx 0.983$.    For both approximations \eqref{graph1} and \eqref{graph2}  the $T_i$ values are  the  same. This result is another confirmation of the consistency of our model.   
\\
Based on the variation of $k$ one can simply explain the  Leidenfrost phenomenon as follows.   If $k>0$, the thermodynamic pressure $p$ is lower in the vapor phase just near  the liquid-vapor interface compared to the pressure $p_0$ near the surface. This results  to a detachment of the liquid film from the surface. On the contrary, if $k<0$, the thermodynamic pressure $p$ is higher,  and   liquid film wets  the surface causing a violent boiling. In fact the whole process is highly non-stationary and cannot be described by the stationary equations. However, our approach gives a reasonable estimation of the Leidenfrost temperature. 
	
\begin{figure}
	\begin{center}
		\includegraphics[width=\linewidth]{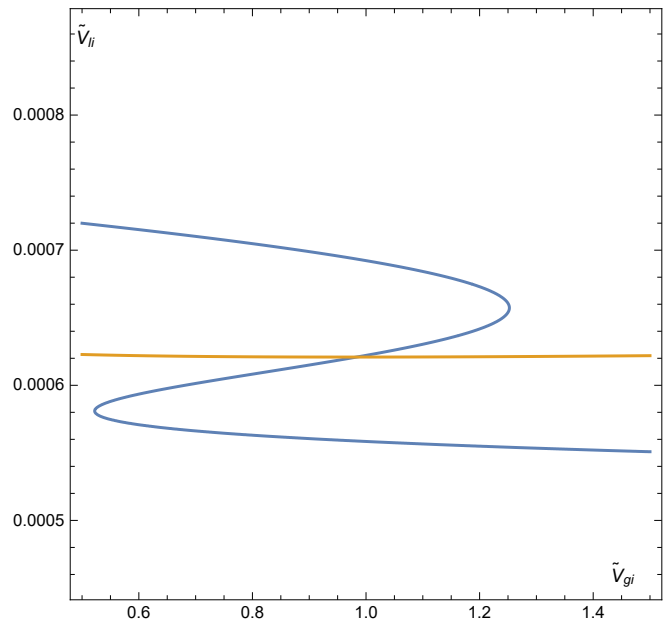}
	\end{center}
	\caption{ Contour--graphs associated with the two first equations of system \eqref{systeme2} are shown in the case $\tilde T_w= 1.23$, when $\tilde T_i$ is eliminated from the third equation. The $Z-shape$ curve corresponds to the Eq. \eqref{systeme2}$_2$, the second curve corresponds to Eq. \eqref{systeme2}$_1$. The horizontal (vertical) axis indicates  dimensionless vapor and liquid specific volumes, respectively.  The curves intersect transversally in a unique  point. These graphs prove  that  the calculated solution  is little sensitive to the approximation  of physical parameters.} \label{Fig.6}
\end{figure}

\section{Conclusion}
For water, we study  the  boiling crisis  phenomenon in the framework of the   internal capillarity  model.\\  A first important result is  the  capillary pressure term $k$  allows us to understand the phenomenon and to determine the Leidenfrost temperature. The boiling crisis corresponds to $k>0$, and Leidenfrost's temperature to $k=0$.   For water, the model
predicts  the Leidenfrost temperature which fairly agrees with experimental
results. 
A second important result is the estimation of  the   temperature  of   liquid--vapor interface of water. It is proved that its value is   below the boiling temperature at atmospheric pressure. This result is also consistent with experimental data on the overall liquid temperature  near the interface.
\\
 In the future, we plan to apply this model to other fluids for which all necessary experimental data are well documented. \\

\textbf {Acknowledgments}: The authors thank D. Brutin and B. Darbois-Texier for pointing out to us useful references, and  the reviewers for a  careful reading of the manuscript and for many helpful inquiries which allowed us to improve our paper.	The authors   are partially supported  by Agence Nationale de la Recherche, France (SNIP ANR--19--ASTR--0016--01).

\appendix
\label{Appendix}
\section{Extra-condition at dynamical liquid-vapor interfaces}\label{AnnexA}

Extra--condition \eqref{additivecond} does not come from  conservation laws. It is  a natural boundary condition coming from Lagrangian formulation of the problem. It already appeared in the  study of discontinuous solutions of dispersive equations  \cite{Gavrilyuk,Gavrilyuk1, Gavrilyuk2}.  To give  a  proof in the one-dimensional case,  we consider a general action functional: 
\begin{equation*}
\mathcal A\{y\}=\int_{ I}{\cal L}(y,y')dx,  
\end{equation*}
$y(x)$ is an  unknown function,
and the integral is taken over a finite interval  {\it I}. The values of  $y(x)$ are fixed at the ends of   interval {\it I}. We are looking for $y(x)$ on which the functional is extremal and we do not assume that $y(x)$ is smooth. 
The variation of  Hamilton's action $\mathcal A$  can be written as:
\begin{equation*}
\begin{array}{l}
\displaystyle\delta  \mathcal A =\int_{I}\left\{\frac{\delta{\cal L}}{\delta y}\,\delta y+\frac{d}{dx}\left(\frac{\partial {\cal L}}{\partial y^\prime}\,\delta y\right)\right\}dx,\\ \\ \displaystyle \quad{\rm with}\quad \frac{\delta{\cal L}}{\delta y}=\frac{\partial {\cal L}}{\partial y}-\frac{d}{dx}\left(\frac{\partial {\cal L}}{\partial y^\prime}\right).
\end{array} 
\end{equation*}
In the case of non--smooth (or "broken") extremal curves, the same Euler--Lagrange equation should be  satisfied  for  each smooth part of the extremal curve:
\begin{equation}
\frac{\delta{\cal L}}{\delta y}=0.
\label{eq1}
\end{equation}
Together with Eq. \eqref{eq1} an  additional   condition should also be satisfied at  the "broken" point: 
\begin{equation}
\left[\,\frac{\partial {\cal L}}{\partial y^\prime}\,\right]=0. 
\label{eq2}
\end{equation}
In the case of capillary fluids,  $\cal{L}$ is quadratic with respect to $y^\prime$ because $\lambda$ is constant. It implies that  $y^\prime$ is  continuous at the \textit{broken} point. Condition \eqref{eq2} is usually called \textit{Weierstrass-Erdmann} condition,  or \textit{corner} condition.   In particular, if  a piecewise $C^2$--solution $y(x)$ is constant on some  interval of $x$, 
but is not constant on a neighboring interval, this solution should have a zero slope at the broken point.

\section{Special cases of    capillary fluid motion}\label{Annex}
\subsection{Isothermal motion}\label{Isothermal motion}
In the case of isothermal stationary motion,
the whole entropy of  domain ${\cal{D}}_t$ corresponding to the bulk (a) and interface (i) is:
\begin{equation}
\int_{{\cal{D}}_t} \rho \eta \, d D= S_0,\label{constraint1}
\end{equation}
where $S_0$ is   constant (independent of time $t$), and $dD$ is the  infinitesimal  volume. Due to constraint \eqref{constraint1}, Hamilton's action is modified into the following: there exists a constant Lagrange multiplier $T_0$ such that the new Lagrangian $\mathcal L$ is associated with   $  \alpha- T_0\, \eta$ which is   the specific free energy at constant temperature. The application of the Hamilton principle yields the same equations of motion where $\alpha$ has to be replaced by $\alpha-T_0 \eta$. Consequently, the specific enthalpy is replaced by the chemical potential $\mu$. The variation of $\eta$ implies $T-T_0=0$.

\subsection{Motion at constant pressure}\label{Motion at constant pressure}

In the case of stationary motion, if  domain ${\cal{D}}_t$ is  an invariant control volume through which the steam  flows, it verifies: 
\begin{equation}
\int_{{\cal{D}}_t} \, dD ={\mathcal V}_0,\label{constraint2}
\end{equation} 
where ${\mathcal V}_0$ is   constant (independent of time $t$).  Due to constraint \eqref{constraint2}, Hamilton's action is modified into the following: there exists a constant Lagrange multiplier $p_0$   such that the new Lagrangian $\mathcal L$ is associated with   ${\cal H}=\alpha + p_0/\rho$,   which is  the specific enthalpy of capillary fluid at constant pressure $p_0$.  Consequently,  in  Subsection \ref{Conservative motion}, in the energy equation, the specific energy should be replaced by  the specific enthalpy at constant pressure $p_0$, and the equation of motion  is unchanged.

\section{Isothermic  oscillations of the vapor density near liquid-vapor interface
	(i)}\label{oscillations}
\begin{figure}[!h]
	\begin{center}
	\includegraphics[width=\linewidth]{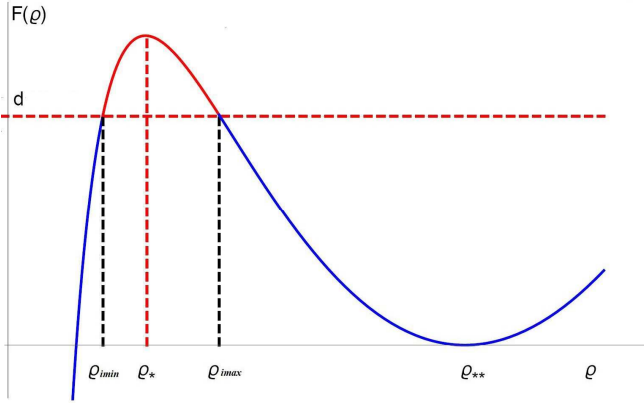}
	\caption{{If $M_i^2<1$, the curve $F(\rho)$ has a local maximum  at $\rho_{\star}\in \left]\;\rho_{imin}, \rho_{imax}\right[$, and a local minimum  at $ \rho_{\star\star}\in \left]\; \rho_{imax}, +\infty\right[$. We recall that $\rho=\rho_{\star\star}$ is  a formal  value of $\rho$ and that the  physical part of the curve is only the {\it red part} of $F(\rho)$ corresponding to oscillations of density between    $\rho_{imin}$ and $\rho_{imax}$.}}
	\label{7}
	\end{center}
\end{figure}
	\begin{figure}[!h]
	\begin{center}
		\includegraphics[width=\linewidth]{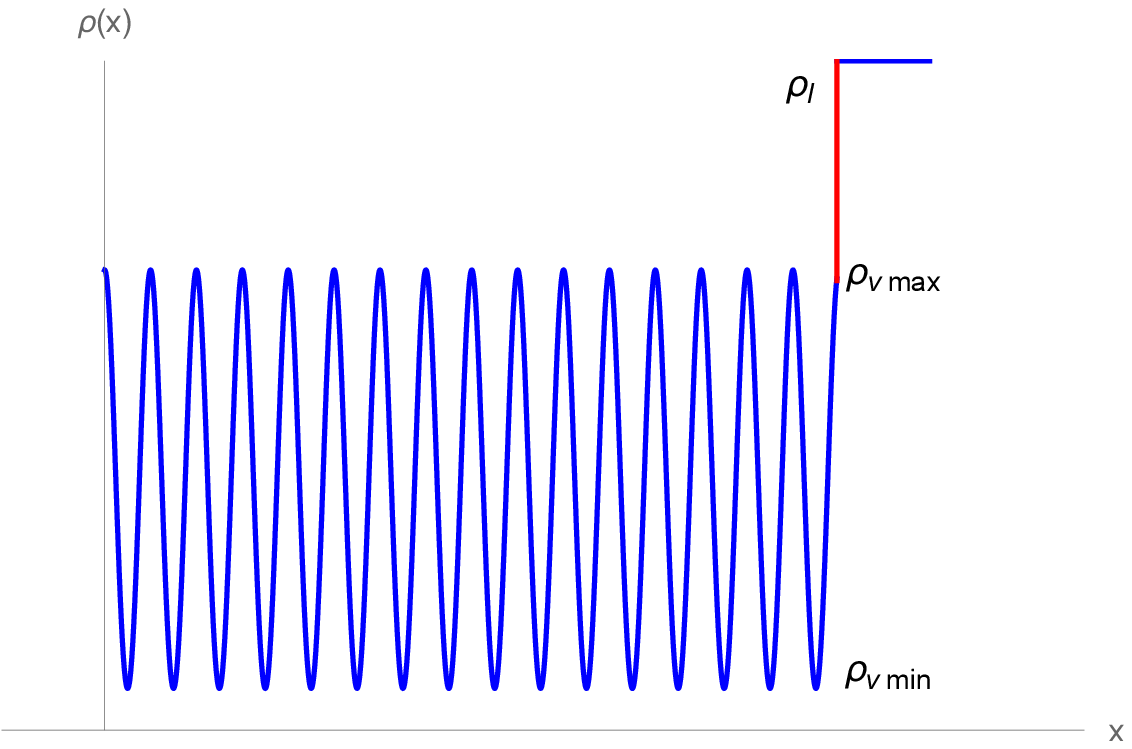}	
		\includegraphics[width=\linewidth]{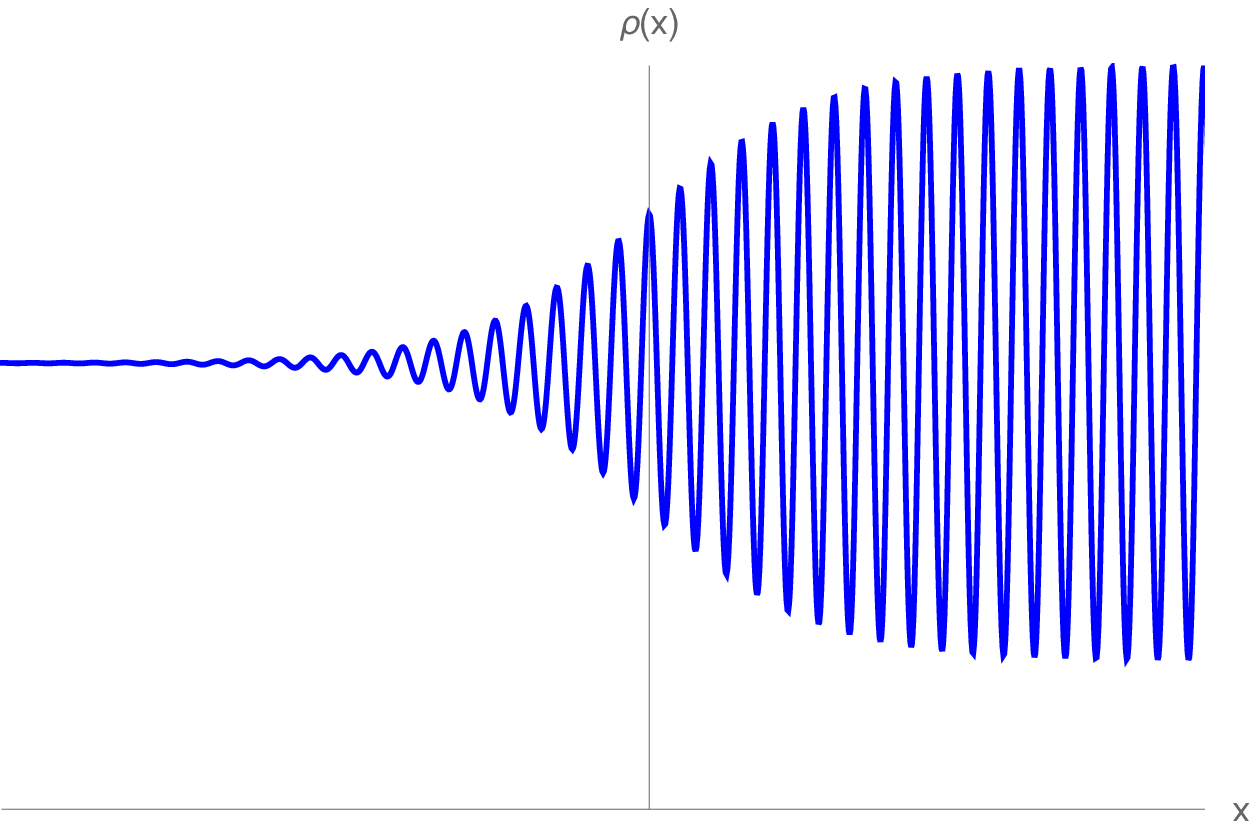}
	\end{center}
	\caption{Upper figure :  The vapor density   oscillates   near the isothermal liquid-vapor interface.  At the interface the density jumps (and  decreases) from     $\rho_{l_i}=1/v_{l_i}$ to $\rho_{g_i}=1/v_{g_i}$. The vapor density   being oscillating  between two extrema $\rho_{imin}$ and $\rho_{imax}$ where $\displaystyle d\rho/dx =0$, we have to choose between these  two values. The jump from $\rho_{l_i}$ to $\rho_{imax}$  has a smaller amplitude compared to that  from $\rho_l$ to $\rho_{imin}$, and hence a smaller energy decrease. Consequently, $\displaystyle {d^2 \rho}/{dx^2}<0$ when  $\rho_{v_{i}}=\rho_{imax}$ and $k=-\lambda\, \rho  \, {d^2 \rho}/{dx^2}>0$. 
		Bottom figure : case of   dissipative vapor flow. The oscillations of vapor density vanish  near the surface boundary layer. }
	\label{8}
\end{figure}

We look for oscillating  stationary vapor flow  in the immediate vicinity  of  interface  (i)  where    the temperature is $T_i$.    The  governing equation  of motion in  the vapor phase is deduced from Eqs \eqref{chemicalpotential} and \eqref{mass1}, and writes in the form: 
\begin{equation*}
\lambda \frac{d^2\rho}{dx^2}=\mu (\rho, T_i)+\frac{q^2}{2\rho^2}+ r,
\label{3}
\end{equation*}
	where  $r$ is a constant of integration.
The   vapor is considered as an ideal gas; we get the potential $\mu$, defined up to an   additive constant which can be included in $r$: 
\begin{equation*}
\mu(\rho, T_i)=c_{T_i}^2\,{\ \rm Log} \,  \rho  ,\quad {\rm where} \quad c_{T_i}^2= {\mathcal R}T_i.
\end{equation*}
Here $c_{T_i}$ denotes the isothermal    sound  velocity of vapor at temperature $T_i$.
To obtain oscillatory solutions, we choose a special value of $r$ replacing it by a new constant $\rho_{\star\star}$ : 
\begin{equation}
\lambda \frac{d^2\rho}{dx^2}=c_{T_i}^2{\ \rm Log}\left(\frac{\rho}{\rho_{\star\star}}\right)+\frac{q^2}{2\rho^2}-\frac{q^2}{2\rho_{\star\star}^2}.
\label{second_derivative}
\end{equation}
Integrating Eq. \eqref{second_derivative}, one obtains:
\begin{equation}
\frac{\lambda}{2} \left(\frac{d\rho}{dx}\right)^2=F(\rho)-d,\quad{\rm where}\quad d={\rm const}.
\label{first_integral}
\end{equation}
with 
\begin{equation*}
F(\rho)=c_{T_i}^2\left(\rho{\ \rm Log}\left(\frac{\rho}{\rho_{\star\star}}\right)-\rho+\rho_{\star\star}\right)-\frac{q^2}{2\rho}\left(1-\frac{\rho}{\rho_{\star\star}}\right)^2. 
\end{equation*} 
By construction,
\begin{equation*}
F(\rho_{\star\star})=0,\quad \frac{dF}{d\rho}\left(\rho_{\star\star} \right)=0.
\end{equation*}
The variation of $F$   for the \textit{Mach number}  $M_i^2= {q^2}/({\rho_{\star\star}\, c_{T_i}})^2 <1$  is shown in Fig.\ref{7} ({In liquid water $\rho_l\simeq 10^3 kg/m^3$. If  boiling--evaporation time of a liquid film with $10^{-2}   $ m   thickness is about $100 \, s$; then $u\simeq 10^{-4} m/s$ and for the liquid, the flow rate  $q\simeq 10^{-1} kg/m^2 .\, s$. In the vapor $\rho_l\simeq 1\,  kg/m^3$ and consequently  $u\simeq 10^{-1} m/s$, and $M_i^2
	\simeq 10^{-7} < 1$}). It has a unique maximum point $\rho_{\star}$ such that $0<\rho_\star<\rho_{\star\star}$. Moreover, $F\rightarrow -\infty$ as $\rho\rightarrow +0$. Hence, for any $d$ such that  $0<d<F(\rho_{\star})$ one has a solution of \eqref{first_integral} oscillating between $\rho_{imin}$ and $\rho_{imax}$, where $F(\rho_{imin})=F(\rho_{imax})=d$  (see Fig. \ref{7}). The solution of \eqref{first_integral} is schematically shown in Fig. \ref{8} 
(on the  high diagram).   The liquid--vapor interface is considered as a discontinuity. So, the density jumps from $\rho_{l_i}$ to  the extreme value of the vapor density (see the extra condition \eqref{additivecond}). Since we have two possible values (minimum and maximum values), the choice has to be done. Obviously, the jump from $\rho_{l_i}$ to $\rho_{imax}$  has a smaller amplitude compared to that  from $\rho_l$ to $\rho_{imin}$, and hence the smallest energy variation. Also, physically, only this choice allows us to obtain `levitation' of the liquid film. 
Such  a stationary periodic solution gives us    an idea  about  a strong density variation near  the interface:   the liquid-vapor interface is endowed with a micro-structure representing a strongly oscillatory region. The vapor region represents   a  transition zone  that begins  with an oscillatory regime  and ends with  a region of   homogeneous density  (see the bottom diagram in  Fig. \ref{8}).

\end{document}